\documentstyle[12pt]{article}
\newcommand{\lesssim}
{\mathrel{\raisebox{-2.8pt}{\mbox{$\stackrel{\textstyle <}{\sim}$}}}}

\newcommand{\address}[1]{\centerline{\small\it #1}}
\newcommand{\draft}{}
\renewcommand{\title}[1]{\begin{center}%
{\large\bf #1}\end{center}\par\bigskip}
\renewcommand{\author}[1]{\centerline{#1}}
\renewcommand{\maketitle}{}
\newcommand{\pacs}[1]{}
\newcommand{\narrowtext}{}

\begin{document}
\draft
\title{From Quantum Dynamics to the Canonical Distribution\\
-- General Picture and a Rigorous Example\footnote{
The first version July 1997.
The final version December 1997.
Archived as cond-mat/9707253.
To appear in Phys. Rev. Lett.
}}
\author{Hal Tasaki$^*$}
\address{Department of Physics, Gakushuin University,
Mejiro, Toshima-ku, Tokyo 171, JAPAN}
\maketitle
\begin{abstract}
Derivation of the canonical (or Boltzmann) distribution
based only on quantum dynamics is discussed.
Consider a closed system which consists of
mutually interacting
subsystem and heat bath, and assume that the whole
system is initially in a pure state
(which can be far from equilibrium)
with small energy fluctuation.
Under the ``hypothesis of equal weights for
eigenstates'', we derive the canonical distribution in the sense that,
at sufficiently large and typical time,
the (instantaneous) quantum mechanical expectation value of an
arbitrary operator of the subsystem is almost equal to the desired
canonical expectation value.
We present a class of examples in which the above derivation
can be rigorously established without any
unproven hypotheses.
\end{abstract}
\narrowtext

It is often said that the principles of equilibrium statistical
physics have not yet been justified.
It is not clear, however, what statement should be
regarded as the ultimate justification.
Recalling the astonishingly universal applicability of equilibrium
statistical physics, it seems likely that there are many independent
routes for justification
which can be equally convincing and
important \cite{note1,notechaos}.
In the present paper, we concentrate on one of the specific
scenarios for obtaining canonical distributions from
quantum dynamics \cite{ref}.

\paragraph*{Outline of the work:}
Let us outline our problem and the main result.
We consider
an isolated quantum mechanical system
which consists of a
subsystem and a heat bath.
The subsystem is described by a Hamiltonian $H_{\rm S}$ which have
arbitrary
nondegenerate eigenvalues $\varepsilon_{1},\ldots,\varepsilon_{n}$.
For convenience we let
$\varepsilon_{j+1}>\varepsilon_{j}$ and $\varepsilon_{1}=0$.
The heat bath is described by a Hamiltonian $H_{\rm B}$ with the
density of states $\rho(B)$.
The inverse temperature of the heat bath at energy
$B$ is given by the
standard formula $\beta(B)=d\log\rho(B)/dB$.
We assume (as usual) $\beta(B)$ is
positive and decreasing in $B$.
The density of states
$\rho(B)$ is arbitrary except for a fine structure that we
will impose on the spectrum of $H_{\rm B}$.

The coupling between the subsystem and the heat bath is given by
a special
Hamiltonian $H'$ which almost conserves the unperturbed energy and
whose magnitude is $\Vert H'\Vert\sim\lambda$.
We assume $\Delta\varepsilon\gg\lambda\gg\Delta B$,
where $\Delta\varepsilon$ is the minimum spacing of the energy levels
of $H_{\rm S}$,
and $\Delta B$ is the maximum spacing of that of $H_{\rm B}$.
These conditions guarantee a
weak coupling between the subsystem and the
bath, as well as macroscopic nature of the bath.
The Hamiltonian of the whole system is
$H
=
H_{\rm S}\otimes{\bf 1}_{\rm B}
+{\bf 1}_{\rm S}\otimes H_{\rm B}
+H'$,
where ${\bf 1}_{\rm S}$ and ${\bf 1}_{\rm B}$ are the identity
operators for the subsystem and the heat bath, respectively.

Suppose that the whole system is initially in a pure
state $\Phi(0)$
which has an energy distribution peaked around (but not strictly
concentrated at) $E$.
It is possible to treat mixed states as well, but such extensions are
not essential.
For an operator $A$ of the subsystem, we denote its
quantum mechanical expectation value at time $t$ as
\begin{equation}
	\langle A\rangle_{t}
	=
	\langle\Phi(t),(A\otimes{\bf 1}_{\rm B})\Phi(t)\rangle,
	\label{At}
\end{equation}
where $\langle\cdot,\cdot\rangle$ stands for the
inner product, and $\Phi(t)=e^{-iHt}\Phi(0)$ is the state at time
$t$.
Note that $\langle\cdots\rangle_{t}$ is a mixed state on the
subsystem.
Our main result is the derivation of the canonical distribution in
the sense that
\begin{equation}
	\langle A\rangle_{t}
	\simeq
	\langle A\rangle^{\rm can}_{\beta(E)}
	\quad
	\mbox{for any $A$,}
	\label{goal}
\end{equation}
holds \cite{note2} for sufficiently large and typical $t$,
where
$\langle A\rangle^{\rm can}_{\beta}
=
{\rm Tr}_{\rm S}[Ae^{-\beta H_{\rm S}}]
/
{\rm Tr}_{\rm S}[e^{-\beta H_{\rm S}}]$
is the canonical expectation value.
We show that (\ref{goal}) holds for rather general systems under
the ``hypothesis of equal weights for eigenstates.''
For a special class of models, we prove (\ref{goal}) rigorously
without any unproven hypotheses.

We note the following points about the present derivation of
the canonical distribution.
(i)~We do not introduce any probability distributions by hand.
(ii)~We do not make use of the microcanonical distribution.
(iii)~We do not perform any time averaging.
(iv)~We do not take any limits such as making
the bath infinitely large or the coupling infinitesimally small.
(v)~Quantum mechanics seems to play essential roles.

In the present paper,
we describe our main results and basic idea of proofs, leaving
details to \cite{me}.
We also briefly discuss possible extension of the present
scenario to more general systems.

\paragraph*{Coupling:}
We diagonalize the (partial) Hamiltonians as
$H_{\rm S}\Psi_{j}=\varepsilon_{j}\Psi_{j}$ with
$j=1,\ldots,n$,
and
$H_{\rm B}\Gamma_{k}=B_{k}\Gamma_{k}$ with
$k=1,\ldots,N$,
where $\Psi_{j}$, $\Gamma_{k}$ are normalized.
We will impose a fine structure on the spectrum
$\{B_{k}\}$ when we discuss our rigorous
results.

When the coupling $H'$ is absent, the total Hamiltonian
$H_{0}
=
H_{\rm S}\otimes{\bf 1}_{\rm B}
+{\bf 1}_{\rm S}\otimes H_{\rm B}$
has eigenstates
$\Theta_{(j,k)}=\Psi_{j}\otimes\Gamma_{k}$
with eigenvalues
$U_{(j,k)}=\varepsilon_{j}+B_{k}$.
We now introduce a new index $\ell=1,\ldots,nN$ for $\Theta$ and $U$.
The index $\ell$ is in a one-to-one correspondence with the original
index $(j,k)$ such that $U_{\ell+1}\ge U_{\ell}$ holds for
$\ell=1,\ldots,nN-1$.
We define the coupling Hamiltonian $H'$ as
\begin{equation}
	\langle\Theta_{\ell},H'\Theta_{\ell'}\rangle
	=
	\cases{
	\lambda/2,&if $|\ell-\ell'|=1$;\cr
	0,&otherwise,\cr
	}
	\label{H'}
\end{equation}
with a constant $\lambda>0$.
The Hamiltonian $H'$ describes scattering processes which almost
conserve the unperturbed energy.

\paragraph*{Eigenstates---main idea:}
Let $\Phi_{E}$ be a normalized eigenstate of the
total Hamiltonian $H=H_{0}+H'$ with energy $E$.
Expanding it as
$\Phi_{E}=\sum_{\ell=0}^{nN}\varphi_{\ell}\Theta_{\ell}$,
the Schr\"{o}dinger equation
$E\Phi_{E}=H\Phi_{E}$ is written as
\begin{equation}
	E\varphi_{\ell}=
	\frac{\lambda}{2}(\varphi_{\ell-1}+\varphi_{\ell+1})
	+U_{\ell}\varphi_{\ell},
	\label{Sch}
\end{equation}
with $\varphi_{0}=\varphi_{nN+1}=0$.
This may be regarded as the Schr\"{o}dinger equation for a single
quantum mechanical ``particle'' on a ``chain'' $\{1,2,\ldots,nN\}$
under the monotone ``potential'' $U_{\ell}$.

The eigenstates of (\ref{Sch}) with a constant $U_{\ell}=U$ are
$\varphi_{\ell}=e^{\pm ik\ell}$
with $k=\cos^{-1}[(E-U)/\lambda]$ if $|E-U|<\lambda$, and
$\varphi_{\ell}=e^{\pm\kappa\ell}$
(or $\varphi_{\ell}=(-1)^{\ell}e^{\pm\kappa\ell}$)
with $\kappa=\cosh^{-1}[|E-U|/\lambda]$ if $E-U>\lambda$
(or $E-U<-\lambda$).
Since our ``potential'' $U_{\ell}$ varies slowly in $\ell$, the
quasi-classical argument \cite{Landau} suggests that generally
$\varphi_{\ell}$ takes appreciable
values in the ``classically
accessible region'' (which consists of $\ell$ such that
$|E-U_{\ell}|\lesssim\lambda$), and is negligible
outside the region.
More precisely we expect that, for general $E$,
we can write
$|\varphi_{\ell}|^{2}\approx f(E-U_{\ell})$
with
a ($E$ dependent) function $f(\tilde{E})$ which is
nonnegligible only for
$|\tilde{E}|\lesssim\lambda$.
In terms of the original index, this reads
\begin{equation}
	|\varphi_{(j,k)}|^{2}\approx f[E-(\varepsilon_{j}+B_{k})].
	\label{f}
\end{equation}

To see consequences of (\ref{f}),
we take an arbitrary operator $A$ of the
subsystem, and denote its matrix elements as
$(A)_{j,j'}=\langle\Psi_{j},A\Psi_{j'}\rangle$.
By using (\ref{f}) and noting that
$\Delta\varepsilon\gg\lambda\gg\Delta B$,
we find
\begin{eqnarray}
	\langle\Phi_{E},(A\otimes{\bf 1}_{\rm B})\Phi_{E}\rangle
	  &=&
	 \sum_{j,j'=1}^{n}\sum_{k=1}^{N}
	 \varphi_{(j,k)}^{*}\varphi_{(j',k)}(A)_{j,j'}
	\nonumber  \\
	 &\simeq&
	 \frac{\sum_{j,k}|\varphi_{(j,k)}|^{2}(A)_{j,j}}
	 {\sum_{j,k}|\varphi_{(j,k)}|^{2}}
	\nonumber \\
	&\simeq&
	 \frac
	 {\sum_{j}\int dB\,\rho(B)f(E-\varepsilon_{j}-B)(A)_{j,j}}
	 {\sum_{j}\int dB\,\rho(B)f(E-\varepsilon_{j}-B)}
	\nonumber  \\
	&\simeq&
	 \frac
	 {\sum_{j}\rho(E-\varepsilon_{j})(A)_{j,j}}
	 {\sum_{j}\rho(E-\varepsilon_{j})}
	\nonumber  \\
	 &\simeq&
	 \langle A\rangle^{\rm can}_{\beta(E)},
	\label{EAE}
\end{eqnarray}
where the final estimate follows by Taylor expanding
$\rho(E-\varepsilon_{j})$ as usual.

The relation (\ref{EAE}) states that the expectation value in an
eigenstate is equal to the desired canonical expectation value.
This is the key estimate in the present work, and the rest of
our results follow from relatively
general (and standard) arguments.
Although we have restricted our discussion to the $H'$ of the form
(\ref{H'}), we expect (and can partially prove \cite{me}) that
the property (\ref{f}) holds for general
eigenstates of systems with more general couplings $H'$.
This may be called \cite{me} the ``hypothesis of equal weights for
eigenstates'', from which we may
get the key estimate (\ref{EAE}) and its consequences.

\paragraph*{Eigenstates---rigorous result:}
We will precisely state the assumption on
$H_{\rm B}$, and describe a rigorous estimate corresponding to
(\ref{EAE}).
We fix an energy unit $\delta>0$ (which may be much smaller than
$\lambda$), a positive integer $R$, and positive integers
$M_{1},M_{2},\ldots,M_{R}$.
We then introduce an integer $L\ge L_{0}$ (where $L_{0}$ is a
constant), which will be made
sufficiently large (but finite)
to realize the situation where the bath is
``large.''
We require for each $r=1,\ldots,R$ that $LM_{r}$ eigenvalues of
$H_{\rm B}$ are distributed in the interval $((r-1)\delta,r\delta)$
with an equal spacing $b_{r}=(LM_{r})^{-1}\delta$.
Thus the density of states of the bath is written as
$\rho(r\delta)=LM_{r}/\delta$.

With this special $H_{\rm B}$, we
can partially control the solution of (\ref{Sch}) and the sums in
(\ref{EAE}) to get the following \cite{proof}.

{\em Lemma} ---
Consider a system where
$H_{\rm B}$ has the above fine structure, and the coupling
$H'$ is given by (\ref{H'}).
We assume $\varepsilon_{j+1}-\varepsilon_{j}\ge4\lambda$ for any $j$.
Let $E$ be an eigenvalue of the whole Hamiltonian $H$ such that
$\varepsilon_{n}+2\lambda\le E\le
B_{\rm max}-2\lambda$,
and $\Phi_{E}$ be the corresponding eigenstates.
Then for any operator $A$ of the subsystem, we have
\begin{equation}
	\left|
	\langle\Phi_{E},(A\otimes{\bf 1}_{\rm B})\Phi_{E}\rangle
	-
	\langle A\rangle_{\beta}^{\rm can}
	\right|
	\le
	\sigma\Vert A\Vert.
	\label{EAER}
\end{equation}
Here
$\sigma=3\beta\lambda+\gamma(\varepsilon_{n})^{2}+cL^{-1/12}$
with
$\beta=\beta(E)=d\log\rho(E)/dE$,
$\gamma=|d\beta(E-\lambda)/dE|$,
$\Vert A\Vert=\max_{j,j'}|(A)_{j,j'}|$,
and $c$ is a constant independent of $L$, $E$, and $A$.

Note that we have $\sigma\ll1$
if
(i)~the coupling is weak (to have $\beta\lambda\ll1$),
(ii)~$\beta(E)$ varies slowly
(to have $\gamma(\varepsilon_{n})^{2}\ll1$), and
(iii)~the level spacing of the bath is small
(to have $cL^{-1/12}\ll1$).
Recall that
these are the standard assumptions regarded necessary to get the
canonical distribution.
We have established the key estimate (\ref{EAE}) rigorously
under reasonable conditions.

\paragraph*{Long-time average:}
Let $\Phi(0)$ be the initial state (of the whole system), and
expand it as
\begin{equation}
	\Phi(0)=\sum_{E}\gamma_{E}\Phi_{E},
		\label{Gamma1}
\end{equation}
where the sum runs over the eigenvalues of $H$.
The state at time $t$ is given by
$\Phi(t)=\sum_{E}e^{-iEt}\gamma_{E}\Phi_{E}$.
Then the quantum mechanical expectation value (\ref{At})
can be written
as
\begin{equation}
	\langle A\rangle_{t}
	=
	\sum_{E,E'}e^{i(E-E')t}(\gamma_{E})^{*}\gamma_{E'}
	\langle\Phi_{E},
	(A\otimes{\bf 1}_{\rm B})\Phi_{E'}\rangle.
	\label{At2}
\end{equation}
Since energy eigenvalues of (\ref{Sch}) are nondegenerate, we find
that the long-time average of $\langle A\rangle_{t}$ becomes
\begin{eqnarray}
	\overline{\langle A\rangle_{t}}
	&=&
	\sum_{E'}|\gamma_{E'}|^{2}
	\langle\Phi_{E'},
	(A\otimes{\bf 1}_{\rm B})\Phi_{E'}\rangle
  \nonumber \\
	&\simeq&
	\sum_{E'}|\gamma_{E'}|^{2}
	\langle A\rangle^{\rm can}_{\beta(E')},
	\label{At3}
\end{eqnarray}
where
$\overline{F(t)}=\lim_{T\to\infty}T^{-1}
\int_{0}^{T}dtF(t)$, and
we used (\ref{EAE}) or (\ref{EAER}) to get the final line.

Further suppose that the initial state $\Phi(0)$ has a small energy
fluctuation in the sense that the coefficients $\gamma_{E'}$ is
nonnegligible only for $E'$ close to some fixed energy $E$.
Then (\ref{At3}) reduces to
\begin{equation}
	\overline{\langle A\rangle_{t}}
	\simeq
	\langle A\rangle^{\rm can}_{\beta(E)},
	\label{LTA}
\end{equation}
which states that the long-time average of the quantum mechanical
expectation value is almost equal to
the desired canonical expectation value \cite{cat}.

An interesting example of an initial state with small energy
fluctuation is
\begin{equation}
	\Phi(0)
	=
	\Psi_{n}\otimes\sum_{k}\alpha_{k}\Gamma_{k},
	\label{Phi2}
\end{equation}
with $\alpha_{k}$ nonnegligible for $k$ in a finite range.
Note that the state (\ref{Phi2}), restricted onto the subsystem, is
very far from equilibrium since $n$ is the index for the highest
energy state of the subsystem.

\paragraph*{Approach to equilibrium:}
We have seen that, for the initial state with small energy
fluctuation, the time-independent part in the right-hand side of
(\ref{At2}) gives the desired canonical expectation value.
The time-dependent part of (\ref{At2}) is a linear combination of many
terms oscillating in $t$ with different frequencies.
It might happen that at sufficiently large and typical (fixed)
 $t$, these
oscillating terms cancel out with each other, and
$\langle A\rangle_{t}$ becomes almost identical to its
time-independent part $\overline{\langle A\rangle_{t}}$.

This naive guess is strengthened by the following simple
estimate of
the variance.
\begin{eqnarray}
	\overline{
	\left(
	\langle A\rangle_{t}
	-
	\overline{\langle A\rangle_{t}}
	\right)^{2}
	}
	&=&
	\overline{(\langle A\rangle_{t})^{2}}
	-
	\left(
	\overline{\langle A\rangle_{t}}
	\right)^{2}
	\nonumber\\
	&=&
	\sum_{E_{1},E_{2},E_{3},E_{4}}
	(\gamma_{E_{1}})^{*}\gamma_{E_{2}}
	(\gamma_{E_{3}})^{*}\gamma_{E_{4}}
	\overline{
	e^{i(E_{1}-E_{2}+E_{3}-E_{4})t}
	}
	\langle E_{1}|A|E_{2}\rangle
	\langle E_{3}|A|E_{4}\rangle
	\nonumber\\
	&&
	-\left(
	\sum_{E}|\gamma_{E}|^{2}
	\langle E|A|E\rangle
	\right)^{2}
	\nonumber\\
	&\le&
	\sum_{E,E'}|\gamma_{E}|^{2}|\gamma_{E'}|^{2}
	\langle E|A|E'\rangle\langle E'|A|E\rangle
	\nonumber\\
	&\le&
	n^{2}\Vert A\Vert^{2}\max_{E}|\gamma_{E}|^{2},
	\label{var}
\end{eqnarray}
where we used the Dirac notation
$\langle E|A|E'\rangle
=\langle\Phi_{E},(A\otimes{\bf 1}_{B})\Phi_{E'}\rangle$,
and used the bound
$\langle E|A^{2}|E\rangle\le n^{2}\Vert A\Vert^{2}$.
We also assumed the non-resonance condition for the the
relevant energy
eigenvalues, i.e., whenever $E_{1}-E_{2}=E_{4}-E_{3}\ne0$ holds for
$E_{i}$ such that $\gamma_{E_{i}}\ne0$, we have
$E_{1}=E_{4}$ and $E_{2}=E_{3}$.
Although we are not able to verify the non-resonance condition for a
particular given model, it is easily proved that the condition is
satisfied for generic models \cite{NRC}.
When a large number of states equally contribute to the expansion
(\ref{Gamma1}), $|\gamma_{E}|^{2}$ and hence
the right-hand side of (\ref{var})
is very small.
This means that $\langle A\rangle_{t}$ usually takes values very
close to its average $\overline{\langle A\rangle_{t}}$.

Following the idea of the Chebyshev's inequality \cite{Feller},
this observation can be made into a rigorous statement.
Instead of writing down the general theorem \cite{me}
we present its consequence for the special
(but interesting) situations with
the initial states (\ref{Phi2}).

{\em Theorem} ---
Suppose that the conditions of the Lemma hold, and the
non-resonance condition is valid
for energy eigenvalues with
$\varepsilon_{n}+2\lambda\le E_{i}\le
B_{\rm max}-2\lambda$.
Fix an arbitrary energy $E$ with
$\varepsilon_{n}+3\lambda\le E\le B_{\rm max}-3\lambda$.
Take an initial state $\Phi(0)$ of the form (\ref{Phi2}), and assume
that  $\alpha_{k}$ is nonvanishing only for $k$ such that
$|E-(B_{k}+\varepsilon_{n})|\le\varepsilon_{n}/2$, and satisfies
$(\alpha_{k})^{2}\le c'(\varepsilon_{n}\rho(E))^{-1}$ with an arbitrary
constant $c'\ge1$ \cite{alpha}.
We let
$\sigma'=3\beta\lambda+3\gamma(\varepsilon_{n})^{2}+cL^{-1/12}
+n^{2}(\lambda/\varepsilon_{n})^{1/3}$, where
$\beta$, $\gamma$ are the same as in the Lemma.
Then there exists a finite $T>0$ and a subset (i.e., a collection of
intervals) ${\cal G}\subset[0,T]$ with the following properties.
a)~For any $t\in{\cal G}$, we have
\begin{equation}
	|\langle A\rangle_{t}
	-
	\langle A\rangle^{\rm can}_{\beta}|
	\le
	\sigma'\Vert A\Vert,
	\label{A-A}
\end{equation}
for any operator $A$ of the subsystem.
b)~If we denote by $\mu({\cal G})$ the total length of the intervals in
$\cal G$, we have
\begin{equation}
	1\ge
	\frac{\mu({\cal G})}{T}
	\ge
	1-3c'\left(\frac{\lambda}{\varepsilon_{n}}\right)^{1/3}.
	\label{main}
\end{equation}

In simpler words, $\cal G$ is the collection of ``good''
time intervals, in which
the quantum mechanical expectation value of {\em any} operator
is equal to
the canonical expectation value within the relative error
$\sigma'$.
The new factor $\lambda/\varepsilon_{n}$ is small if the coupling is
weak.
Therefore under the physical
conditions (i)-(iii) stated below the Lemma,
both $\sigma'$ and $3c'(\lambda/\varepsilon_{n})^{1/3}$
in (\ref{main}) are small.
Then (\ref{main}) says that the ``good'' intervals
essentially cover the whole interval $[0,T]$.
We thus have obtained the desired (\ref{goal}) for
most $t$ within the time interval $0\le t\le T$.
Recall that the subsystem is far from equilibrium at $t=0$.
{\em
By using only quantum dynamics and the assumptions about the initial
state,
we have established an approach to the desired canonical distribution
from a highly
nonequilibrium state.}
Although our theorem is proved only for systems with particular
fine structure and special $H'$,
the essential physics is contained in
the ``hypothesis of equal weights for eigenstates'' and
the non-resonance condition (or related weaker conditions).
We expect the same scenario to work for much more general systems.

Our result is not strong enough to clarify how
$\langle A\rangle_{t}$ approaches (or deviates from)
its equilibrium value.
Note that we can never expect a perfect decay to equilibrium since
$\langle A\rangle_{t}$ is quasiperiodic in $t$.
In a long run, $\langle A\rangle_{t}$ deviates from the equilibrium
value infinitely often.
Our theorem does guarantee that such deviations are indeed very rare
for a weak (but finite) coupling.
Unfortunately we do not have any meaningful estimate for $T$.

\paragraph*{Irreversibility:}
One might question if our theorem implies the existence of
an ``irreversible'' time evolution towards equilibrium.
We believe the answer is affirmative, but we have to be careful about
this delicate issue.
We first stress that, exactly as in classical cases \cite{note1},
whether we see irreversibility or not depends on the choice of
physical quantities that we observe.
We should clearly see irreversibility when
(there is irreversibility and) we observe
quantities which have small fluctuation in the equilibrium (and
preferably in the initial state too) \cite{macro}.

To get an illustrative example, we suppose $n$ is large, and set
$A$ to be the projection onto the state $\Psi_{n}$.
$A$ has eigenvalues 0 and 1.
In the initial state $\Phi(0)$ of the form (\ref{Phi2}), an
observation of $A$ does not disturb the state and one gets a definite
result 1.
We then wait for a sufficiently long time (required by the theorem),
and once again
observe $A$ at time $t$.
If this $t$ does not belong to the exceptional ``bad'' intervals, the
theorem guarantees that
$\langle A\rangle_{t}\simeq\langle A\rangle^{\rm can}_{\beta}
=O(1/n)$.
This does {\em not} mean we observe a value of $O(1/n)$, but means
we observe the definite value 0 with probability $1-O(1/n)$ which is
essentially 1 for large enough $n$.
This is what we mean by (macroscopic) irreversibility
\cite{note1,qm}.

\bigskip
It is a pleasure to thank
Jean Bricmont,
Takashi Hara,
Yasuhiro Hatsugai,
Tetsuya Hattori,
Kazuo Kitahara,
Mahito Kohmoto,
Tohru Koma,
Joel Lebowitz,
Seiji Miyashita,
Tohru Miyazawa,
Yoshi Oono,
and
Tooru Taniguchi
for stimulating discussions on various related topics.


\begin{thebibliography}{10}

\bibitem[*]{email}
Electronic address:
hal.tasaki@gakushuin.ac.jp

\bibitem{note1}
We believe that the (still)
widespread scenario of obtaining microcanonical distributions from the
ergodicity (in the strongest form) can hardly be justified.
For careful discussions about these points, see, for examples,
J. L.Lebowitz, Physica A {\bf 194}, 1 (1993);
Physics Today, Sept. 1993, 32;
in {\em 25 Years of Non-Equilibrium Statistical Physics}
eds. J. J. Brey et al., Lecture Notes in Physics 445,
(Springer 1995).
Y. Oono,
`Statistical Thermodynamics'
in Encyclopedia of Polymer Science and
Engineering, 2nd Edition (John Wiley) vol.15., 614-625 (1989).
J. Bricmont, Physicalia Magazine 17, 159-208
(1995), reprinted in {\em The Flight from Science and Reason}
 (New
York Academy of Sciences, New York, 1996),
archived as chao-dyn/9603009.

\bibitem{notechaos}
We also feel too much emphasis on the possible roles of
chaos (either classical or quantum)
in the foundation of statistical physics can be misleading.
But for attempts to derive
statistical distributions using results and conjectures from ``quantum
chaos'', see
M. Srednick, Phys. Rev. E {\bf 50}, 888 (1994);
J. Phys. A {\bf 29} L75 (1996).

\bibitem{ref}
For numerical experiments which confirm similar scenarios, see
R. V. Jensen and R. Shanker,
Phys. Rev. Lett. {\bf 54}, 1879 (1985);
K. Saito, S. Takesue and S. Miyashita,
J. Phys. Soc. Jpn. {\bf 65}, 1243 (1996),
Phys. Rev. E {\bf 54}, 2404 (1996).
There is a vast literature for situations where an
(infinitely large) heat bath is put
into an equilibrium distribution (by hand), and time evolution of
attached subsystem is discussed.

\bibitem{note2}
Note that (\ref{goal}) is true for {\em any} operators.
Therefore when one ``observes'' $A$,
one does not see the expectation value itself, but
finds fluctuation according to the prediction of the
canonical distribution.
The origin of the fluctuation
may be traced back to the probabilistic interpretation in the
Copenhagen spirit.
It is correct to say that (at least in the present way of introducing
the canonical distribution and within the standard interpretation of
quantum mechanics) there is no intrinsic distinction between quantum
fluctuation and thermal fluctuation.
(Our results themselves are free from any specific
interpretations of quantum mechanics.)
We stress, however, that the present one is not at all
the only reasonable way of introducing
probability.  See, for examples, \cite{note1}.

\bibitem{me}
H. Tasaki, to be published.
An unpublished note which (only) contains all the technical
details is available as cond-mat/9707255 (or from the author).

\bibitem{Landau}
L. D. Landau and E. M. Lifschitz,
Quantum Mechanics (Non-relativistic Theory) (Pregamon, 1977).

\bibitem{proof}
Even with the fine structure on $H_{\rm B}$,
we are not able to control the
global solution of (\ref{Sch}), and the actual proof is rather
involved.
In the classically accessible region, we divide the whole range of
$\ell$ into small intervals, and prove in each interval
that $\varphi_{\ell}$ has the form suggested
by the (discrete version of) quasi-classical analysis \cite{Landau};
$\varphi_{\ell}\simeq
A\cos(\sum_{\ell'=\ell_{0}}^{\ell}k_{\ell'}+\theta)/
\sqrt{\sin k_{\ell}}$
with $k_{\ell}=\cos^{-1}[(E-U_{\ell})/\lambda]$.
The estimate of the sums in (\ref{EAE}) is nontrivial because
$\varphi_{\ell}$ oscillates.
Near the turning points, we approximate (\ref{Sch}) by a continuous
equation and estimate $\varphi_{\ell}$ using Bessel's functions.
Details can be found in \cite{me}.

\bibitem{cat}
The assumption of small energy fluctuation is {\em physically
necessary} no matter how one ``derives'' the canonical distribution.
Without the condition, we end up with a Schr\"{o}dinger's cat type
state where states with different temperatures are superposed.

\bibitem{NRC}
Let us restrict ourselves to the non-resonance condition for
eigenvalues with $E\ge \varepsilon_{n}+2\lambda$
(which is all we need).
Suppose that we have a model which violates the non-resonance condition.
Then by slightly shifting $B_{k}$ in the lowest two bands $(0,\delta)$
and $(\delta,2\delta)$,
we get a model which satisfies the condition \cite{me}.

\bibitem{Feller}
W. Feller,
{\em An Introduction to Probability Theory and Its Applications I},
(Wiley, 1968)

\bibitem{alpha}
We took the allowed energy width equal to
$\varepsilon_{n}$ only to make formulae simpler.
We can prove the corresponding estimates for any energy width.
For $c'\simeq1$ the upper bound for $(\alpha_{k})^{2}$
means that all the
allowed basis states in (\ref{Phi2}) contribute almost equally.

\bibitem{macro}
Macroscopic observables in macroscopic systems automatically have such
properties.

\bibitem{qm}
We are {\em not} trying to get irreversibility from
disturbances by observations.
We believe it highly misleading to expect any positive roles of
observations in (macroscopic) irreversibility and stochastic
behavior.

\end{thebibliography}
\end{document}